\documentclass[copyright,creativecommons]{eptcs}
\providecommand{\event}{PLACES 2024} 

\usepackage{listings}
\usepackage{listings-rust}
\usepackage{fancyvrb}
\usepackage{multirow}
\usepackage{csquotes}
\usepackage{graphicx}
\usepackage{microtype}
\usepackage{hyperref}
\usepackage{blindtext}
\usepackage{calc}
\usepackage{enumitem}
\usepackage{amsmath}
\usepackage{amssymb}
\usepackage{tikz}
\usetikzlibrary{positioning}
\usetikzlibrary{arrows.meta}
\usetikzlibrary{patterns}

\newcommand\tcp{\textsc{tcp}}
\newcommand\tcpstate[1]{\textsc{#1}}

\definecolor{bluetypes}{HTML}{004488}
\newcommand{\tycolour}[1]{\textcolor{bluetypes}{#1}}
\newcommand{\tyfontstyle}[1]{\texttt{{#1}}}
\newcommand{\ty}[1]{\tycolour{\tyfontstyle{#1}}}

\definecolor{redroles}{HTML}{CC3311}
\newcommand{\rolecolour}[1]{\textcolor{redroles}{#1}}
\newcommand{\role}[1]{\texttt{\rolecolour{#1}}}

\definecolor{greenmessages}{HTML}{117733}
\newcommand{\messagecolour}[1]{\textcolor{greenmessages}{#1}}
\newcommand{\stmessage}[1]{\texttt{\messagecolour{#1}}}

\newcommand{\lbl}[1]{\texttt{#1}}

\lstnewenvironment{rust}[1][]
{\lstset{captionpos=b,style=colouredRust,language=Rust,#1}}
{}

\newcommand{\Rinline}[1][]{\lstinline[style=colouredRust,language=Rust,#1]}

\hypersetup{hidelinks}

\renewcommand{\textsc}{\uppercase}

\title{Session Types for the Transport Layer: Towards an Implementation of TCP\footnote{Supported in part by the UK EPSRC grants EP/X027309/1 and EP/S036075/1.}}

\author{Samuel Cavoj
\email{samuel@cavoj.net}
\institute{University of Glasgow}
\and
Ivan Nikitin 
\email{ivan@niktivan.org}
\institute{University of Glasgow}
\and
Colin Perkins 
\email{csp@csperkins.org}
\institute{University of Glasgow}
\and
Ornela Dardha 
\email{ornela.dardha@glasgow.ac.uk}
\institute{University of Glasgow}
}

\begin{document}

\newcommand{\authorrunning}{S. Cavoj, I. Nikitin, C. Perkins, O. Dardha}
\newcommand{\titlerunning}{Session types for TCP}
\maketitle

\begin{abstract}
Session types are a typing discipline used to formally describe communication-driven applications with the aim of fewer errors and easier debugging later into the life cycle of the software.
Protocols at the transport layer such as TCP, UDP, and QUIC underpin most of the communication on the modern Internet and affect billions of end-users.
The transport layer has different requirements and constraints compared to the application layer resulting in different requirements for verification.
Despite this, to our best knowledge, no work shows the application of session types at the transport layer.
In this work, we discuss how multiparty session types (MPST) can be applied to implement the TCP protocol.
We develop an MPST-based implementation of a subset of a TCP server in Rust and test its interoperability against the Linux TCP stack.
Our results highlight the differences in assumptions between session type theory and the way transport layer protocols are usually implemented. This work is the first step towards bringing session types into the transport layer.

\end{abstract}

\section{Introduction}

Session types~\cite{honda-1998} are a typing discipline for communication protocols. They can describe the sequence of
messages exchanged between participants over a communication channel and can be used to verify that
the protocol is implemented correctly or has certain desirable properties. 
Further, session types can be realised within programming languages and used to type-check the implementation of a protocol against a session type definition, with type errors indicating inconsistencies between implementation and the session type.
Session types have been an active area of research since the beginning of the 1990s \cite{honda-1998} and have been implemented in a
number of programming languages including C \cite{mpst-c}, Java \cite{st-java} and Rust
\cite{st-rust-1, Kokke_2019} and other programming languages \cite{Fowler16, KokkeD21, NgY16, Padovani17, PucellaT08}.

Network protocols that are part of the Internet Protocol suite (\textsc{tcp/ip}) are the foundation of the
Internet. 
They are responsible for interoperability between different devices, operating systems, and
applications. 
To ensure that different implementations of the same protocol are compatible, they
must adhere to a technical specification which, in the case of Internet protocols, is defined in a
series of documents, known as \textsc{rfc}s \cite{rfc8700}, developed by the Internet Engineering Task Force (IETF). 
Specifically, the latest version of the \tcp{} protocol specification is defined in \textsc{rfc}~9293 \cite{rfc9293}.

The IETF follows a consensus-based process when developing standards \cite{rfc2026,rfc7282}, with protocol specifications being developed in working group meetings and on mailing lists over a multi-year period. The resulting RFCs are written primarily in English prose, allowing the documents to be used in the consensus-building process, but the natural language can be ambiguous and unclear and this can lead to inconsistent and non-conforming implementations. \cite{MBDP2020,MBDP2021,paxson1997automated}. 
In this sense, ensuring the correctness of Internet protocols is vital.
Developing formalised models of the protocols described in RFCs is one way to achieve this.
Session types are one such modelling technique that has not previously been explored for transport-layer protocols, such as TCP.

In this paper, we implement a core subset of the \tcp{} protocol in the Rust programming language and use session types to describe the network operations. Session types are encoded into native Rust types
and the type checker is used to verify that the implementation follows the session type specification.
In this way, the Rust compiler verifies that the implementation of the protocol
is correct in terms of the types of messages exchanged and the order in which they are exchanged,
i.e., that it follows the declared session type, for a session type model describing a synchronous subset of TCP.
Additionally, session types are used to describe the application interface, so we can verify that the
application uses the \tcp{} implementation correctly.

Our contributions are as follows:
\begin{enumerate}
    \item \textbf{Session Types Libraries.} We develop\footnote{Our session type library and TCP implementation is available at \url{https://github.com/sammko/tcpst2}} the libraries required for encoding the session type model into native Rust types in an ergonomic fashion (\S\ref{st:rust:defining}).
    \item \textbf{Implementation.} We implement a subset of the \tcp{} protocol \cite{rfc9293}, including key aspects of both the user/\tcp{} interface and the \tcp{}/lower-level interface, in Rust
        while adhering to the session type model. This is done in a way such that the
        Rust compiler can detect deviation from the session type (\S\ref{st:rust:usage}).
    \item \textbf{Testing.} We test our implementation against a real \tcp{} stack (\S\ref{ch:evaluation}).
\end{enumerate}



The remainder of this paper is structured as follows. Section \ref{st} briefly reviews the multiparty session type model we use. Section \ref{tcp} outlines key properties of TCP and its state machine. Section \ref{ch:implementation} describes our session typed implementation of TCP in Rust. Section \ref{ch:evaluation} evaluates the correctness of our implementation. Finally, Section \ref{related} reviews related work and concludes.

\section{Session types}\label{st}

Session types~\cite{honda-1998} describe communication among participants
in a distributed system in terms of the types and order of messages that are exchanged.
A single session type describes the sequence of messages sent or received from the perspective of one
of the participants. The theory of session types was later extended to multiparty
session types (\textsc{mpst}) which can describe protocols between any number of
participants \cite{honda2008multiparty}. 

In this paper, the bottom-up multiparty session type approach~\cite{less-is-more} is used to describe \tcp{}.
An example of a simple ping-pong protocol using this approach is demonstrated in Equation~\ref{eqn:mpst_example_simple}.
When type-checking any type using the bottom-up approach, we must additionally choose a \textit{safety invariant}.
Safety invariants are parameters associated with the properties a protocol may demonstrate during runtime, such as deadlock-freedom and liveness. Each safety invariant is accompanied by specific typing rules (not presented here) that guarantee the maintenance of the corresponding invariant. If the protocol successfully type-checks with the instantiation of the safety invariant, it will manifest the property represented by the invariant during its runtime.

\begin{equation}
    \label{eqn:mpst_example_simple}
    \begin{array}{rl}
        \tycolour{\ensuremath{\Gamma_1}} \; = \; & \texttt{s}[\role{a}]:\role{b}\ \ensuremath{\oplus}\ \lbl{$l_1$}(\stmessage{ping})\ .\ \role{b}\ \ensuremath{\:\&\:}\ \lbl{$l_3$}(\stmessage{pong})\ .\ \ty{end}, \cr
        & \texttt{s}[\role{b}]:\role{a}\ \ensuremath{\:\&\:}\ \lbl{$l_2$}(\stmessage{ping})\ .\ \role{a}\ \ensuremath{\oplus}\ \lbl{$l_1$}(\stmessage{pong})\ .\ \ty{end}
    \end{array}
\end{equation}

The implications of this approach are that global types and the concept of duality are not used.
Instead of duality, the compatibility invariant is used to check that actions are dual between the given types.
However, a protocol can still be described using session types even if safety does not hold.

\section{Transmission Control Protocol (TCP)}\label{tcp}


The \tcp{} transport is layered on top of the datagram service provided by the Internet Protocol (IP). The IP layer provides an unreliable, best-effort, datagram service, where packets may be lost, duplicated, delayed, or re-ordered in transit.
TCP segments, sent within IP packets, contain sequence numbers and acknowledgements such that, upon detection of a lost packet, either triggered by a timer expiration or receipt of a triple-duplicate acknowledgement, the sender can re-transmit the lost segment.

\begin{figure}[t]
\centering
\includegraphics[width=0.75\textwidth]{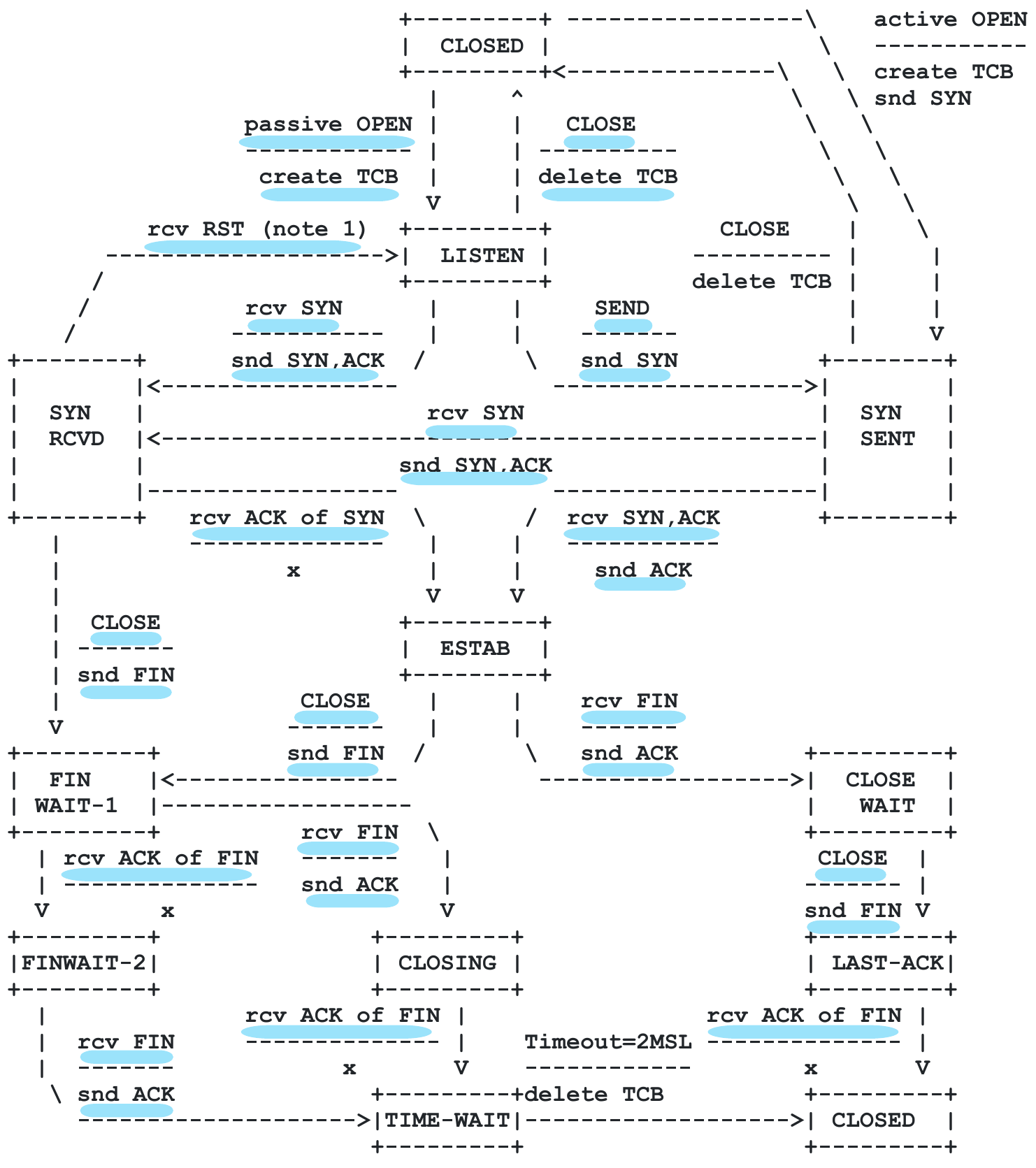}
\caption{The state transition diagram of TCP in RFC9293~\cite{rfc9293}. We annotate the diagram with the messages and transitions modelled in our implementation. Note that we do not model timeouts as part of the type system, hence, the TIME-WAIT to CLOSED transition is not implemented using session types. Additionally, we do not implement the active OPEN case of the handshake for simplicity (as this would not demonstrate any new modelling or implementation techniques), this is however possible using our implementation.}
\label{app:tcp_diagram}
\end{figure}

\tcp{} is usually used in a client-server manner, but also supports a rarely used simultaneous open mode with peer-to-peer connections.
In the context of this paper, we assume client-server usage, with one side being a passive server listening for incoming connections, while the other is an active client initiating the connection.
We describe the operation of the TCP state machine below and provide a diagram of the TCP state transitions in Figure~\ref{app:tcp_diagram}.

The establishment of a reliable connection between two network devices is facilitated by the TCP three-way handshake. It commences with the initiation of a connection with the client sending a TCP segment with the SYN (synchronise) bit set in the header and containing the client's initial sequence number. 
The server responds with a segment with the SYN and ACK bits set, acknowledging the client's initial sequence number and providing the initial sequence number the server will use.
Finally, the client confirms the establishment of the connection by sending a segment with the ACK (acknowledge) bit set. 
This sequence ensures both sides agree on their initial sequence numbers and confirm their willingness to communicate.

TCP uses a sliding window algorithm to manage data transmission by sending segments with sequence numbers. The window size determines the number of unacknowledged segments in transit. The receiver discards unacceptable segments falling outside the expected sequence range, leading to retransmission by the sender. Acknowledgements are sent upon receiving new data, indicating the next expected contiguous sequence number. TCP handles packet loss or reordering at the IP layer by detecting duplicate acknowledgements; a triple-duplicate acknowledgement triggers retransmission. Additionally, TCP utilises a retransmission timeout (RTO) mechanism, dynamically adjusted based on network conditions. TCP buffers play a crucial role on both the sender and receiver sides, with the send buffer holding outgoing segments awaiting acknowledgement and the receive buffer storing incoming segments yet to be delivered to the application.

The TCP closing handshake, another three-way handshake involving packets with the FIN (finish) and ACK (acknowledge) bits signifies the end of a connection. The initial party sends a FIN packet, followed by an acknowledgement from the other party, culminating in a reciprocal FIN-ACK exchange. The final step includes an acknowledgement from the original sender, leading to the TIME-WAIT state. This state ensures a reliable closure, allowing the handling of delayed or duplicate IP packets before concluding the connection.

\section{Implementation}\label{ch:implementation}

We implement
the basic functionality of the \tcp{} server protocol while
modelling both the network and the application interface using session types. 
Note that more information on the implementation can be found in the Appendix.
Under the \textit{less
is more} formalisation of multiparty session types~\cite{less-is-more}, the roles we are
considering the following:

\begin{description}[leftmargin=!, labelwidth=\widthof{\bfseries Server System}]
    \item[Server User] The server application using the \tcp{} protocol.
    \item[Server System] The \tcp{} implementation.
    \item[Client System] The \tcp{} implementation on the other end of the network.
\end{description}

The channel between the Server System and the Client System represents the
network.
The messages exchanged between the Server User and the Server System are a
formalisation of the user/\tcp{} (i.e., application programming)
interface and do not pass over the network. The system call interfaces, representing the user and the system (in this paper simulated through threads), each have a session type which prescribes their behaviour relative to the other roles. The Client System role has no associated
session type in our implementation as it is assumed to be another host on the Internet and not part
of our program.

\subsection{Defining session types}\label{st:rust:defining}

The basic building blocks of our implementation are the generic structs \Rinline{OfferOne},
\Rinline{OfferTwo}, \Rinline{SelectOne}, and \Rinline{SelectTwo}. All of these implement the trait
\Rinline{Action} which represents a general session type. The type parameters of the structs
encode the role the action is performed with respect to, the types of messages exchanged, and
the continuation of the session. In addition, the \Rinline{End} struct is also an \Rinline{Action}
and represents the end session type.

The \Rinline{OfferTwo} struct has five type parameters. The
first is the peer role, and the next two are the types of messages exchanged in either of the two
branches of the offer and the final two parameters are the session types of the continuations of the
two branches.

\begin{rust}
pub struct OfferTwo<R, M1, M2, A1, A2>
where R: Role, M1: Message, M2: Message,
    A1: Action, A2: Action,
{
    phantom: PhantomData<(R, M1, M2, A1, A2)>,
}
\end{rust}
The \Rinline{OfferTwo} struct, as a way of encoding a session type construct in Rust, has type parameters but contains no data. The \Rinline{PhantomData}-typed field contained within the struct is a zero-sized marker type that simulates a field of the given type to support the Rust type checker.\footnote{\url{https://doc.rust-lang.org/nomicon/phantom-data.html}}

The \Rinline{SelectTwo} struct has the same type parameters and is also a zero-sized type.
Finally, the non-branching actions \Rinline{OfferOne} and \Rinline{SelectOne}
have only three type parameters: the peer role, the message type, and the continuation type,
but are otherwise analogous.

To define a session type one can define a type alias for the root action of the session. For example
a simple session type for a client-server interaction could be defined as follows:
\begin{rust}
type ServerSt = OfferOne<Client, Request, SelectOne<Client, Response, End>>;
type ClientSt = SelectOne<Server, Request, OfferOne<Server, Response, End>>;
\end{rust}

This basic syntax, however, quickly becomes unwieldy when defining more complex session types. To
address this, we have implemented a macro which converts a more readable syntax into the full
definition of the type. Rust's \Rinline|macro_rules!| mechanism is powerful enough to allow us to
define a syntax which attempts to mimic the mathematical notation. The macro is called \Rinline|St!|
and the \Rinline{ServerSt} type from the above example could be re-written as follows:
\begin{rust}
type ServerSt = St![(Client & Request).(Client + Response).end]
\end{rust}

The macro is recursive and supports arbitrary nesting of offers and selections. The full definition
can be found in the \texttt{st\_macros.rs} file of the source code.

\subsection{Multi-way \texttt{Offer} branching}\label{st:rust:multiway}

As Rust does not support variadic generic types, we are not aware of a way to implement a generic
\Rinline{Offer} type which would support a variable number of branches. Hence we implement
\Rinline{OfferOne} and \Rinline{OfferTwo} as separate constructs with some repetition in the
corresponding infrastructure such as the \texttt{offer\_one}, \texttt{offer\_two} and similar
selection methods. These are described in~\S\ref{st:rust:usage}.

However, support for more than two branches is required in practice. A simple way to do this is to implement \Rinline{OfferThree}, \Rinline{OfferFour}, \dots, in the same way, along with the code supporting this. This leads to more code duplication, but does not increase complexity, and the usage is straightforward.

As an alternative, to avoid duplication, we chose a nesting approach where a branching of arity \(N\) is transformed into a two-way
branching between the first case and an \(N-1\) branching of the other cases.\footnote{
    Naturally, it would be better to split into halves instead, reducing the expansion depth from
    \(\mathcal{O}(N)\) to \(\mathcal{O}(\log N)\) but this is more difficult to implement and
    provides little practical benefit in all but the most extreme branching cases.
} This is recursively expanded until it finally results in a tree of two-way forks, where each
\textit{left} branch represents a single case from the original \(N\). All \textit{right} branches
except the bottom-most one lead to a virtual node which was not present in the original type.

\subsection{Recursive session types}\label{st:rust:recursion}

Type aliases in Rust cannot be recursive. The reason for this is that a type alias does not create a
new type and is merely another name for the same type. For instance, defining a type alias
\Rinline|type A = X<A>| is not allowed because the expansion would be infinite -- the name
\Rinline{X<A>} would expand to \Rinline{X<X<A>>}, etc. However, we somehow need to represent
recursive session types.

Fortunately, this is not difficult to circumvent. Whereas type
\textit{aliases} cannot be recursive, there is no such restriction for types themselves, as long as
the size of the type is finite. As such, types which contain a recursive cycle with no indirection
are not allowed as the size of the type is infinite. But inserting indirection into the cycle (such
as a reference \Rinline|&T| or \Rinline{Box<T>}) resolves this problem since the size of a reference
does not depend on the size of the target type \Rinline{T}.

\subsection{Using session types}\label{st:rust:usage}

A channel provides methods to send and receive messages which consume a corresponding session type
and return the continuation. The type of the channel is generic over the roles between which it
exists and the method signatures ensure that they can be only called with an appropriate session
type instance and message. Consider a channel of type \Rinline|Channel<R1, R2>| which we
define as the endpoint belonging to role \Rinline{R1}, i.e. it can send to or receive from
\Rinline{R2}. Then its \texttt{select\_one} method could have the following signature:

\begin{rust}
fn select_one<M, A>(&mut self, _o: SelectOne<R2, M, A>, message: M) -> A
where M: Message, A: Action;
\end{rust}

It is generic over the message type, but it has to match the one prescribed by the provided session
typed \textit{token}. The role \Rinline{R2} is already bound by the channel type.
The token is moved into this function, so the owner cannot re-use it.
The continuation type from the token is instantiated and returned to the caller for further
operations. And, of course, the message is transmitted over the underlying transport the nature of
which is not restricted by this abstraction. The only requirement is that the \Rinline{Message} trait
can be converted to a representation that the channel can process, which is the reason for the trait
in the first place.

The implementation of the \Rinline{offer} methods is slightly more involved. Once a message is
received from the underlying transport we must determine which branch of the offer to take and
convert it to the appropriate message type. We outsource the decision to a function we receive as an
argument called the \textit{picker}. We find that in our particular use case, having the capability
to differentiate branches based on external context is necessary. 
This allows us to distinguish the receipt of an expected packet from the error condition when an unexpected packet is received

\subsection{Establishing a Connection}\label{tco:impl:establishing}
A TCP connection is established via a three-way handshake as described in Section \ref{tcp}.
We define the \Rinline{ServerSystemSessionType} to describe creation of the server socket (receipt of \Rinline{Open} from the server user), creating the internal state (the ``TCB''; \S\ref{ap:impl:three-way}), waiting for a SYN from the client, and generating the SYN-ACK segment, corresponding to the transition through the LISTEN state of Figure \ref{app:tcp_diagram} into the SYN RCVD state:
\begin{rust}
pub type ServerSystemSessionType = St![
    (RoleServerUser & Open).
    (RoleServerUser + TcbCreated).
    (RoleClientSystem & Syn).
    (RoleClientSystem + SynAck).
    ServerSystemSynRcvd
];
\end{rust}

The \Rinline{ServerSystemSynRcvd} type describes the SYN RCVD state, with branches indicating the transition to the ESTAB state in \Rinline{ServerSystemCommLoop} if the received ACK is acceptable or closing the connection if not.
\begin{rust}
Rec!(pub ServerSystemSynRcvd, [
    (RoleClientSystem & {
        Ack. // acceptable (i.e., matches the SYN-ACK sent)
            (RoleServerUser + Connected).
            ServerSystemCommLoop,
        Ack. // unacceptable
            (RoleClientSystem + {
                Ack.ServerSystemSynRcvd, Rst.(RoleServerUser + Close).end
            })
     })
]);
\end{rust}
The implementation of three-way handshake is further described in Appendix \ref{ap:impl:three-way}.

\subsection{Data Transmission and Re-transmission}\label{tcp:impl:retransmission}

When a TCP segment goes unacknowledged for a certain amount of time, it is retransmitted.
There are two implementation choices that could be made here: incorporate timeouts into the type system, or leave them out and instead signal session type transitions using external timeouts.
The session type theory we are using does not have a notion of timeouts, nor does any session type work containing timeouts~\cite{BarwellSY022, BocchiMVY19, magpi} have the ability to model the operations needed for \tcp{} timeouts.
Hence, we opt to emulate timeouts by introducing a virtual message type and adding it as another branch to the offer session type. 
In this branch, we continue with a select operation, retransmitting an
\textsc{ack} message and then recursively receiving the next message. The
\Rinline{offer} method on the network channel now accepts another argument, specifying the timeout
duration or \Rinline{None} if no timeout is desired. If the retransmission queue is empty, no timeout
should be employed as we run into an issue if it expires -- the session type requires a segment to
be sent, but there is nothing to send.
Further details around data transmission are in Appendix \ref{ap:impl:exchanging-data}.

\subsection{Closing the connection}\label{tcp:impl:closing}

Closing a \tcp{} connection is a two-step process usually combined into a three-way handshake, as shown in the lower half of Figure \ref{app:tcp_diagram}. 
Each direction of the stream can be closed
independently by sending a segment with the \textsc{fin} bit set. The Server System session type
describes receiving a \textsc{fin} first and then deciding to close eventually, after
allowing the user to send more data using the \Rinline{ServerSystemCloseWait} session type:
\begin{rust}
Rec!(pub ServerSystemCloseWait, [
    (RoleServerUser & {
        Data.
            (RoleClientSystem + Ack).
            (RoleClientSystem & Ack /* empty ack */).
            ServerSystemCloseWait,
        Close.
            (RoleClientSystem + FinAck).
            (RoleClientSystem & Ack).
            end
    })
]);
\end{rust}

The case where the server closes first is handled by the \Rinline{ServerSystemFinWait1}
type:
\begin{rust}
pub type ServerSystemFinWait1 = St![
    (RoleClientSystem & {
        Ack. // ACK of FIN
            ServerSystemFinWait2,
        FinAck. // FIN and ACK of our FIN at the same time
            (RoleClientSystem + Ack).
            end
    })];
\end{rust}
The branch in the \Rinline{ServerSystemFinWait1} type represents the ways in which the closing handshake can proceed after sending a FIN to close the connection and entering into the FINWAIT-1 state (see Figure~\ref{app:tcp_diagram}): either a segment containing an ACK is received causing the system to transition to FINWAIT-2, waiting for a segment containing a FIN indicating that the peer has also finished; or a segment with both FIN and ACK is received causing the final ACK to be sent and terminating the connection via the implied CLOSING and TIME-WAIT states. 
The \Rinline{ServerSystemFinWait2} implementation is analogous, but elided due to space constraints.


Finally, in a full \tcp{} implementation, a \enquote{simultaneous close} situation can occur where
both peers decide to close at the same time. This is not handled by our implementation as it is rarely used and does not fit with the
call-and-response style of interaction we model -- there is no
opportunity for the server to decide to close while waiting for the client.

\section{Evaluation}\label{ch:evaluation}

To evaluate our Server System component, we have implemented a simple echo server in the Server User. 
Every piece of data it receives from the system is split into lines, each line is reversed and then sent back.

The functionality of the server tested is as follows:

\begin{description}
    \item[Establishing a connection] by running netcat and connecting to the server.
    \item[Exchanging data with the client] by typing in messages manually.
    \item[Initiating connection close] by sending an empty line to the server. The server user has
        been programmed to close the connection if an empty line is received.
    \item[Responding to connection close]
        by typing \texttt{\^{}C} which causes netcat to close the socket and therefore send a
        \textsc{fin} to the server.
    \item[Correctly handling a FIN-ACK response to a FIN] by piping an empty line
        immediately followed by \textsc{eof} to netcat. In this situation netcat sends the empty
        line but does not shutdown the socket immediately. Instead it waits for the server to send
        a \textsc{fin-ack} and then sends a \textsc{fin-ack} in response.
\end{description}

We tested our \tcp{} implementation primarily against the Linux kernel \tcp{} stack, running our
program and connecting to it using a Linux user-space \tcp{} client (netcat). 
We have used Scapy~\cite{scapy}, a packet manipulation framework, to emulate a misbehaving \tcp{} client or network and evaluate the behaviour of
our server in response to this. This included sending packets with invalid sequence numbers, invalid
acknowledgement numbers, spurious retransmission or overlapping segments.
Finally, we have tested our implementation against the Linux kernel \tcp{} stack with the addition of
simulated network errors using the \texttt{netem} module to introduce packet loss, delay and reordering. 
Our test script configures the \texttt{TCP\_NODELAY} option on the socket and sends messages in a loop with a small delay between
them. This ensures that the client sends a lot of small packets to observe the effect of packet loss and reordering. 
The received data was then compared to the expected output.
In all cases, we utilised a packet sniffer to monitor the communication.

All of the presented test cases were found to be handled correctly provided the server is in the \tcpstate{established} state. 
During the opening three-way handshake, after the initial \textsc{syn} segment, handling is robust as well.
We found that the server can handle packet loss and reordering errors and the connection can recover once the
impairments are lifted. 
The server does not cache out of order segments in the receive window affecting performance, since these segments will need to be re-transmitted, but not correctness. This is a limitation inherent in using synchronous session types to model an synchronous protocol that permits reordering and packet loss, and suggests future work to extend the modelling approach.

\section{Related Work and Conclusion}\label{related}\label{limit}

Network protocols have been used as examples for various session type theories.
The main protocols used as a demonstration in many works are SMTP~\cite{BMVY19, BFS21, KouzapasDPG16, KouzapasDPG18, LNY22, LM15, LM16} and POP3~\cite{GVR03, ThiemannSts04}.
Both of these protocols are application-layer protocols.
Due to this, models of SMTP and POP3 can assume the guarantees provided by the underlying transport layer protocol -- in most cases this is TCP.
Specifically, any faults, retransmissions and packet re-orderings are handled by the transport layer.
In addition to this, these works do implement or model the protocols exactly from the specification with some works only implementing SMTP partially.
The specific challenges presented by the network link are not considered and the communication channel is considered only in an abstract manner.
This also means that, unlike our implementation of TCP, the works are not shown to connect or work with existing protocol stacks, such as the kernel.
To our best knowledge, ours is the first work to consider the implementation of transport layer protocols from their specification using session types.

In this paper, we have modelled \tcp{} of the Internet protocol suite \cite{rfc9293} using MPST \cite{less-is-more} and implemented a proof of concept in the
Rust programming language, leveraging the Rust type system and borrow checker to verify that the
implementation complies with the session type.
We have successfully tested our implementation using manual
testing against the Linux kernel \tcp{} stack as well as manually constructed \tcp{} segments.
In future work, we aim to address limitations of our implementation such as a lack of timeouts in the type system and the synchronous nature of our implementation.
We additionally aim to model important aspects of the protocol such as congestion control in the future.

\bibliographystyle{eptcs}
\bibliography{main}

\appendix
\section{Appendix}

\subsection{Three-way handshake}\label{ap:impl:three-way}

%

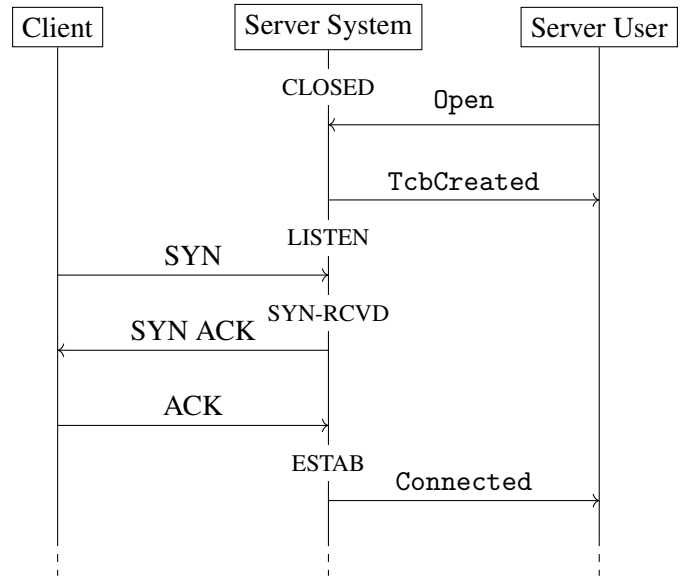
\begin{figure} 
    \centering
    \begin{tikzpicture}[on grid]
        \node[draw, minimum width=2] (client) {Client};
        \node[draw, minimum width=2, right=3.6cm of client] (server) {Server System};
        \node[draw, minimum width=2, right=3.6cm of server] (user) {Server User};

        \foreach \i in {0,...,16} {
                \coordinate (client-\i) at ([yshift=-(\i+0.6)*0.5cm]client);
                \coordinate (server-\i) at ([yshift=-(\i+0.6)*0.5cm]server);
                \coordinate (user-\i) at ([yshift=-(\i+0.6)*0.5cm]user);
            }

        \draw[-] (client.south) -- (client-13) edge[dashed] (client-14);
        \draw[-] (server.south) -- (server-13) edge[dashed] (server-14);
        \draw[-] (user.south) -- (user-13) edge[dashed] (user-14);

        \draw[<-] (server-2) -- node[above] {\texttt{Open}} (user-2);
        \draw[->] (server-4) -- node[above] {\texttt{TcbCreated}} (user-4);
        \draw[->] (client-6) -- node[above] {SYN} (server-6);
        \draw[<-] (client-8) -- node[above] {SYN ACK} (server-8);
        \draw[->] (client-10) -- node[above] {ACK} (server-10);
        \draw[->] (server-12) -- node[above] {\texttt{Connected}} (user-12);

        \node[font=\footnotesize, fill=white] at (server-1) {\tcpstate{closed}};
        \node[font=\footnotesize, fill=white] at (server-5) {\tcpstate{listen}};
        \node[font=\footnotesize, fill=white] at (server-7) {\tcpstate{syn-rcvd}};
        \node[font=\footnotesize, fill=white] at (server-11) {\tcpstate{estab}};
    \end{tikzpicture}
    \caption{TCP three-way Handshake with all roles.}
    \label{fig:impl-tcp-handshake}
\end{figure}

The session type and the \tcp{} state machine are initiated in the \tcpstate{closed} state:
\begin{rust}
let st = ServerSystemSessionType::new();
let tcp = TcpClosed::new();
\end{rust}
The \textit{user} role calls the \Rinline{Open} method and a \textsc{tcb} is created. 
This message is received using \Rinline|offerone| by the system role:
\begin{rust}
let (_open, st) = system_user_channel.offer_one(st);
\end{rust}
The system now transitions to the \tcpstate{listen} state, waiting for a connection establishment to initiate, and sends a \Rinline{TcbCreated} in response:
\begin{rust}
let tcp: TcpListen = tcp.open(LocalAddr { /* ... */ });
let st = system_user_channel.select_one(st, TcbCreated(()));
\end{rust}
The next steps are to wait for a \textsc{syn} segment from the network and respond with a \textsc{syn ack} segment. 
Once a \textsc{syn} segment is received we transition to the \tcpstate{syn-rcvd} state:
\begin{rust}
let (addr, syn, st) = net_channel.offer_one_with_addr(st, &tcp);

let (mut tcp /* Tcp<SynRcvd> */, synack) = tcp.recv_syn(addr, &syn);
let mut syn_rcvd = net_channel.select_one(st, addr, synack);
\end{rust}
The recursive \Rinline{SynRcvd} session type handles unacceptable acknowledgements of \textsc{syn ack} segments which need to be responded to with an \textsc{ack} with the potential of a connection reset:
\begin{rust}
let (mut tcp, st) = loop {
    let st = syn_rcvd.inner();
    let tcp_for_picker = tcp.for_picker();
\end{rust}

The \Rinline|offer_two_filtered| method can now be called on the network channel. 
This method takes the session type, a picker function, and a channel filter:
\begin{rust}
    match net_channel.offer_two_filtered(
        st,
        |packet| match tcp_for_picker.acceptable(&packet) {
            ReactionInner::Acceptable(_, _) => Branch::Left(
                packet.into()),
            _ => Branch::Right(packet.into()),
        },
        &tcp,
    ) {
\end{rust}
Note that the picker determines which branch to take based on the \tcp{} state machine.

The left branch of the session type corresponds to an acceptable \textsc{ack} segment:
\begin{rust}
        Branch::Left((acceptable, st)) => {
            let tcp: Tcp<Established> = tcp
                .recv_ack(&acceptable)
                .empty_acceptable()
                .expect("First ACK must be empty");
            break (tcp, st);
        }
\end{rust}
However, if the \textsc{ack} is not acceptable, the right branch is taken -- an \textsc{ack} is either sent back and the system waits for another \textsc{ack}:
\begin{rust}
        Branch::Right((unacceptable, st)) => {
            let remote_addr = tcp.remote_addr();
            match tcp.recv_ack(&unacceptable) {
                Reaction::NotAcceptable(tcp2, Some(response)) => {
                    let st = net_channel.select_left(
                        st, tcp2.remote_addr(), response);
                    syn_rcvd = st;
                    tcp = tcp2;
                    continue;
                }
\end{rust}
Alternatively, an \textsc{rst}, notifying the user that the connection is being reset:
\begin{rust}
                Reaction::Reset(Some(rst)) => {
                    let st = net_channel.select_right(
                        st, remote_addr, rst);
                    let end = system_user_channel.select_one(
                        st, Close(()));
                    net_channel.close(end);
                    system_user_channel.close(end);
                    return;
                }
\end{rust}
Finally, once out of the loop, the implementation is in the \tcpstate{estab} state and the system notifies the user that the connection is established:
\begin{rust}
let mut recursive = system_user_channel.select_one(st, Connected(()));
info!("established");
\end{rust}
This concludes the implementation of the three-way handshake.

\subsection{Exchanging data}\label{ap:impl:exchanging-data}

The main loop of the implementation waits to receive a segment (using an \Rinline{Offer}
session type) and branches based on their type and whether it is
acceptable or not.

\begin{rust}
Rec!(pub ServerSystemCommLoop, [
    (RoleClientSystem & {
\end{rust}

\begin{description}
    \item[Acceptable with payload] where an acceptable segment is received and there is data present:
\begin{rust}
Ack.
    (RoleClientSystem + Ack /* empty */).
    (RoleServerUser + Data).
    (RoleServerUser & {
        Data.
            (RoleClientSystem + Ack /* with data */).
            ServerSystemCommLoop,
        Close.
            (RoleClientSystem + FinAck).
            ServerSystemFinWait1
    }),
        \end{rust}
        
Initially, data acknowledgement is accomplished using an empty \textsc{ack} segment. The data contained within the \textsc{ack} segment is then transmitted to the server user within a message of type \Rinline{Data}. The user has the option to respond by sending back a message, leading to the transmission of an \textsc{ack} with payload. Alternatively, the user may choose to initiate the closure of the connection, resulting in the transmission of a \textsc{fin ack}.

    \item[Acceptable without payload] In the case where these segments are acknowledgements of previously sent segments, we pass the \textsc{ack} to the \tcp{} state machine to update the state and update the retransmission queue:
        \begin{rust}
Ack.ServerSystemCommLoop,
        \end{rust}

    \item[FIN ACK] The peer has initiated closing the connection. In this case,
        the \tcp{} state machine will transition from \tcpstate{established} to the
        \tcpstate{close-wait} state. Note that due to the absence of timeouts in the type system, the timeout in the close-wait state is implemented outside the session typed state machine.
        \begin{rust}
FinAck.
    (RoleClientSystem + Ack /* we ACK the FIN */).
    (RoleServerUser + Close).
    ServerSystemCloseWait,
        \end{rust}

    \item[Unacceptable] A segment where the sequence numbers are not
        acceptable has been received. The server will respond with an \textsc{ack} which serves to inform
        the peer about the current receive window start and length~\cite{rfc9293}.
        \begin{rust}
Ack.
    (RoleClientSystem + Ack).
    ServerSystemCommLoop,
\end{rust}
\end{description}
\begin{rust}
    })
]);
\end{rust}
\end{document}